\documentclass[aps,epsf,subfigure,twocolumn]{revtex4}

\usepackage{graphicx}% Include figure files
\usepackage{dcolumn}% Align table columns on decimal point
\usepackage{bm}% bold math
\usepackage{amsmath}
\usepackage{subfigure}
\usepackage{color,psfrag,epsfig}

\begin{document}

\newcommand{\bk}{{\bf k}}
\newcommand{\bc}{\begin{center}}
\newcommand{\ec}{\end{center}}
\newcommand{\mub}{{\mu_{\rm B}}}
\newcommand{\sD}{{\scriptscriptstyle D}}
\newcommand{\sF}{{\scriptscriptstyle F}}
\newcommand{\sCF}{{\scriptscriptstyle \mathrm{CF}}}
\newcommand{\sH}{{\scriptscriptstyle H}}
\newcommand{\sAL}{{\scriptscriptstyle \mathrm{AL}}}
\newcommand{\sMT}{{\scriptscriptstyle \mathrm{MT}}}
\newcommand{\sT}{{\scriptscriptstyle T}}
\newcommand{\up}{{\mid \uparrow \rangle}}
\newcommand{\down}{{\mid \downarrow \rangle}}
\newcommand{\upsp}{{\mid \uparrow_s \rangle}}
\newcommand{\downsp}{{\mid \downarrow_s \rangle}}
\newcommand{\upsone}{{\mid \uparrow_{s-1} \rangle}}
\newcommand{\downsone}{{\mid \downarrow_{s-1} \rangle}}
\newcommand{\upt}{{ \langle \uparrow \mid}}
\newcommand{\downt}{{\langle \downarrow \mid}}
\newcommand{\bbar}{{\mid \uparrow, 7/2 \rangle}}
\newcommand{\abar}{{\mid \downarrow, 7/2 \rangle}}
\renewcommand{\a}{{\mid \uparrow, -7/2 \rangle}}
\renewcommand{\b}{{\mid \downarrow, -7/2 \rangle}}
\newcommand{\plus}{{\mid + \rangle}}
\newcommand{\minus}{{\mid - \rangle}}
\newcommand{\psio}{{\mid \psi_o \rangle}}
\newcommand{\psis}{{\mid \psi \rangle}}
\newcommand{\bpsio}{{\langle \psi_o \mid}}
\newcommand{\barpsi}{{\mid \psi' \rangle}}
\newcommand{\barpsio}{{\mid \bar{\psi_o} \rangle}}
\newcommand{\ex}{{\mid \Gamma_2^l \rangle}}
\newcommand{\LH}{{{\rm LiHoF_4}}}
\newcommand{\LHx}{{{\rm LiHo_xY_{1-x}F_4}}}
\newcommand{\de}{{{\delta E}}}
\newcommand{\Ht}{{{H_t}}}
\newcommand{\sL}{{\cal{L}}}
\newcommand{\bfe}{{\bf{e}}}
\newcommand{\bfn}{{\bf{n}}}
\newcommand{\bfR}{{\bf{R}}}

\topmargin=-10mm

\title{What are the interactions in quantum glasses?}

\author{M. Schechter$^1$ and P. C. E. Stamp$^{1,2}$}

\address{$^1$Department of Physics and Astronomy, University of
British Columbia, Vancouver, British Columbia, Canada V6T 1Z1 \\
$^2$Pacific Institute for Theoretical Physics, University of British
Columbia, Vancouver B.C., Canada V6T 1Z1.}

\begin{abstract}

The form of the low-temperature interactions between defects in
neutral glasses is reconsidered. We analyse the case where the
defects can be modelled either as simple 2-level tunneling systems,
or tunneling rotational impurities. The coupling to strain fields is
determined up to 2nd order in the displacement field. It is shown
that the linear coupling generates not only the usual $1/r^3$
Ising-like interaction between the rotational tunneling defect
modes, which cause them to freeze around a temperature $T_G$, but
also a random field term. At lower temperatures the inversion
symmetric tunneling modes are still active - however the coupling of
these to the frozen rotational modes, now via the 2nd-order coupling
to phonons, generates another random field term acting on the
inversion symmetric modes (as well as shorter-range $1/r^5$
interactions between them). Detailed expressions for all these
couplings are given.

\end{abstract}

\maketitle

%%%%%%%%%%%%%%%%%%%%%%%%%%%%%%%%%%%%%%%%%%%%%%%%%%%%%%%%%%%%%%%%%%%%%%%%%%%%%%%%%%%%%%%%%%%%%%%%%%%
%%%%%%%%%%%%%%%%%%%%%%%%%%%%%%%%%%%%%%%%%%%%%%%%%%%%%%%%%%%%%%%%%%%%%%%%%%%%%%%%%%%%%%%%%%%%%%%%%%%
\section{Introduction}
 \label{sec:intro}
%%%%%%%%%%%%%%%%%%%%%%%%%%%%%%%%%%%%%%%%%%%%%%%%%%%%%%%%%%%%%%%%%%%%%%%%%%%%%%%%%%%%%%%%%%%%%%%%%%%
%%%%%%%%%%%%%%%%%%%%%%%%%%%%%%%%%%%%%%%%%%%%%%%%%%%%%%%%%%%%%%%%%%%%%%%%%%%%%%%%%%%%%%%%%%%%%%%%%%%

One of the most subtle and peculiar problems in condensed matter
physics concerns the nature of the "glass" state, and of the
associated glass transition. This problem (described by P. W.
Anderson\cite{And95} in 1995 as "the deepest and most interesting
problem in solid-state theory") concerns the overwhelmingly dominant
component of the physical world as we experience it, ie.,
non-conducting solids that are not ordered in regular crystalline
arrays. In fact the glass problem actually involves two separate
features. One is the remarkable universality displayed in the
low-$T$ quantum properties\cite{YL88,PLT02}, and the other is the
high-$T$ behaviour shown in the vicinity of the glass transition
itself\cite{EAN96}.

At first glance it seems implausible that these two features of
glasses could be related - they occur at very different energy
scales. Elsewhere we argue that there may be an interesting
connection, which depends on certain novel features of the
interactions in these systems. The purpose of the present paper is
to investigate the form of these interactions in some detail. We
derive a number of new interaction terms, which are presented in the
form of two new effective Hamiltonians for neutral glasses, one
valid for higher temperatures, the other in the low-$T$ limit.

We begin with a brief introductory review of the physics of neutral
glasses, particularly in the low $T$ quantum regime. We then derive
the form of the defect-phonon interaction terms (section
\ref{sec:bareH}), including both a direct linear coupling to the
lattice displacement field, and a coupling to the gradient of this
field. We then calculate, in sections \ref{sec:dumb} and
\ref{sec:dumb+}, the effective coupling between defects induced by
these interactions. It is shown in section \ref{sec:dumb} that the
linear coupling not only gives the well-known Ising coupling between
the rotational tunneling modes, but also a random field acting on
these modes. However there is no such linear coupling between
phonons and the 'inversion symmetric defects' (ones which are
symmetric with respect to inversion about the local lattice site).
In section \ref{sec:dumb+} we calculate the coupling of these
defects to gradients of the phonon field, and show that this
coupling produces another weaker coupling to a random field
generated by the now frozen rotational modes, as well as a
short-range coupling between the inversion symmetric tunneling
defects.

The main point of the present paper is to find the correct
quantitative form of the effective Hamiltonian for these systems,
after integrating out the phonons. We end up with an effective
Hamiltonian at high $T$ which involves only the rotational defects,
and then another quite different low-$T$ Hamiltonian which describes
a set of tunneling inversion symmetric defects, coupled to each
other and to the random field generated by the frozen rotational
modes; this includes a number of new terms. It turns out that these
results have very interesting implications for the physics of
glasses, which are discussed in detail in Ref.\cite{SS06}.

%%%%%%%%%%%%%%%%%%%%%%%%%%%%%%%%%%%%%%%%%%%%%%%%%%%%%%%%%%%%%%%%%%%%%%%%%%%%%%%%%%%%%%%%%%%
\subsection{Universalities in the low-T quantum state}
 \label{sec:lowT}
%%%%%%%%%%%%%%%%%%%%%%%%%%%%%%%%%%%%%%%%%%%%%%%%%%%%%%%%%%%%%%%%%%%%%%%%%%%%%%%%%%%%%%%%%%%

The first glass conundrum, strongly emphasized by
Leggett\cite{YL88}, is the apparent universality in the low-T
properties of a huge variety of disordered systems below a
temperature $T_Q \sim 1-3~$K, regardless of the amount of disorder.
The most striking universalities are seen in

(i) the Q-factor $Q(\omega,T)$ for torsional oscillations of the
system, which is related to the phonon mean free path $l(\omega,T)$
and the phonon thermal wavelength $\lambda (\omega ,T)$ by $Q=2 \pi
l/ \lambda$. One finds\cite{PLT02} that $Q(\omega,T)$ shows a
pronounced plateau for $T<T_Q$ (down to a lower temperature which
decreases with $\omega$), and that the value $Q=Q_0 \sim 600$ varies
over only a factor $\sim 2-3$ between many different materials, even
though the intrinsic disorder (measured by e.g., defect
concentration x) may vary over several orders of magnitude.

(ii) The "Berret-Meissner ratio" $c_{\perp}/c_{\parallel}$ between
longitudinal and transverse sound velocities in the same low-T
regime. A remarkable linear relationship is found\cite{BM88} between
$c_{\perp}$ and $c_{\parallel}$, for a variety of materials
including amorphous oxide, semiconducting, polymer, metallic, and
electrolyte glasses, in which $c_{\parallel}$ varies by a factor of
$5$.

These are only the most striking of the low T universalities - there
are others\cite{PLT02,BM88}. In recent years the experimental groups
of Osheroff\cite{LNRO03,LO03}, Hunklinger\cite{NFHE04}, and
Enss\cite{WEL+95} have pushed experiments to very low temperatures
[$ \sim O(1$ mK)], and found a host of interesting new results,
including intrinsic dipolar "hole-digging" in the many-body density
of states\cite{LNRO03}, and a remarkable spin coherence
phenomenon\cite{NFHE04,WEL+95} which comes from nuclear quadrupolar
interactions in neutral glasses. The dynamical hole-digging persists
to the lowest temperatures, giving ever sharper features in the
density of states; it is associated with non-exponential relaxation
and 'aging' behaviour of the dielectric constant of the system. It
would be of interest to continue these measurements well into the
${\rm \mu}K$ regime, if possible.

Although we are clearly dealing here with a resonant tunneling
phenomenon\cite{BNOK98}, qualitatively similar to that in tunneling
spin systems\cite{PS98,Wer01} the glass problem apparently involves
cooperative tunneling of at least coupled pairs of tunneling
systems\cite{LO03,BNOK98}, and the resulting low-T "universal" state
apparently involves some fundamental new physics. Although a number
of theoretical scenarios have been proposed to describe this
universal physics\cite{PLT02,BNOK98,LW07,Par94}, most of which argue
that it must involve strong coupling between the relevant low-$T$
modes, there is no complete consensus at the present time. We note
in passing that although the universalities occur in the same
temperature range as the well-known regularities\cite{HR86} in
thermal and transport properties in glasses (such as the specific
heat $C_V(T) \sim AT$, or the thermal conductivity $K(T) \sim
BT^2$), these latter can all be understood in terms of the
well-known picture\cite{AHV72} of non-interacting two-level
systems(TLSs). On the other hand the dynamics of dipolar
hole-digging certainly requires interactions for its explanation,
between whatever modes are exhibiting low-$T$ quantum fluctuations,
whether these be pairs of TLSs\cite{BNOK98} or some more complicated
set of modes\cite{YL88,LW07}.

%%%%%%%%%%%%%%%%%%%%%%%%%%%%%%%%%%%%%%%%%%%%%%%%%%%%%%%%%%%%%%%%%%%%%%%%%%%%%%%%%%%%%%%%%%%%
\subsection{The high-T glass transition}
 \label{sec:highT}
%%%%%%%%%%%%%%%%%%%%%%%%%%%%%%%%%%%%%%%%%%%%%%%%%%%%%%%%%%%%%%%%%%%%%%%%%%%%%%%%%%%%%%%%%%%%

In most glasses there is actually a glass transition at a
temperature $T_G$ much higher than the crossover temperature $T_Q$
to the universal quantum regime (typically $T_G/T_Q \sim 40$). There
are several universal features of this transition as
well\cite{EAN96}, amongst which one may single out (i) the
characteristic range of relaxation times $\tau$ in the system, and
their characteristic T dependence, summarized in the Vogel-Fulcher
scaling law, which shows that the value of $T_G$ we use depends on
the timescale in interest; (ii) the "entropy crisis", in the range
$T_K < T < T_G$, were $T_K$ is the Kauzman temperature, and where
one finds a supercooled glass entropy lower than that of the
crystalline solid; and (iii) the existence of highly non-exponential
relaxation, and characteristic memory and aging effects, in the
vicinity of the glass transition (as noted above, these effects are
also found in the low-T quantum regime, both in neutral glasses
where universality is seen\cite{LO03} and in electronic
glasses\cite{Ova}).

At first glance there seems to be no obvious relation between this
high-T behaviour, which is characterised by thermally-activated
processes of great complexity, and the low-T behaviour. A number of
attempts have been made in the last couple of years to give a
general theory of the glass transition\cite{LW07,MY06,Lan06}. Two of
them\cite{MY06,Lan06} made no connection to the low-T regime,
instead concentrating on the vast number of thermally-activated
processes coming into play near $T_G$. The Moore-Yeo theory also
makes a very interesting connection between the critical behaviour
of supercooled liquids near $T_G$ and Ising spin glasses in a
magnetic field.

However, there are several arguments that indicate that there may be
a connection between the physics below $T_Q$ and that near $T_G$.
The first two are experimental. It was already noticed by Berret and
Meissner\cite{BM88} that the low-T phonon relaxation time
$\tau_{min}$ shows systematic connection to the values of $T_G$
across the whole range of glasses, with $\tau_{min} \propto
T_G^{2.5}$. We have already remarked on the rough proportionality
$T_G/T_Q \sim 40$; and indeed one can argue that the coupling
$\gamma$ between phonons and the low-T tunneling entities (whatever
they may be) is intrinsically related to both $T_G$ and the phonon
velocities\cite{LW07}.

These observations suggest that there may be some kind of unified
theoretical framework which could describe both the high and low-T
properties of glasses. Such a theory would not only be of major
interest (answering the question posed by Anderson\cite{And95}) but
it would also clearly give us a new blueprint for theories of other
complex systems. Such a framework has actually been proposed very
recently by Lubchenko and Wolynes\cite{LW07}. In this theory the
basic objects are "tunneling centers" which comprise $\sim 200$
atomic units, and which can be used to describe both the low-T
dynamics and the dynamics near $T_G$. We note that these "tunneling
entities" are very different from the TLSs that have been used to
describe many of the low-T
experiments\cite{LO03,NFHE04,WEL+95,BNOK98,HR86}, although it is
argued that their behavior will be quite similar\cite{LW07}.

%%%%%%%%%%%%%%%%%%%%%%%%%%%%%%%%%%%%%%%%%%%%%%%%%%%%%%%%%%%%%%%%%%%%%%%%%%%%%%%%%%%%%%%%%
\subsection{Nature of interactions in glasses}
 \label{sec:intro-int}
%%%%%%%%%%%%%%%%%%%%%%%%%%%%%%%%%%%%%%%%%%%%%%%%%%%%%%%%%%%%%%%%%%%%%%%%%%%%%%%%%%%%%%%%%

We now come to the question to be addressed by this paper. One might
think, in view of the generality of the phenomena discussed above,
that they ought to be independent of the detailed nature of the
interactions between tunneling entities in the low-T regime.
Actually it is widely assumed in the glass literature that it is
enough to include only dipolar strain-mediated interaction, with the
addition of electric dipolar interactions where necessary. This
assumption rests on a number of microscopic calculations done over
the years, both for interacting
TLSs\cite{BNOK98,JL75,BH77,KFAA78,KS89}, and for the more
complicated systems of discrete rotators that exists in
orientational glasses\cite{Mic86,MR85,LM94,GRS90}. These
calculations (particularly those for orientational glasses) are very
lengthy, and for this reason have not been attempted with any
generality except by a few authors. Nevertheless, the conclusion has
been that the effective interaction between $2$ tunneling entities a
distance $R_{12}$ apart is
\begin{equation}
H_{int} \sim g(a_0/R_{12})^3 \label{dipint}
\end{equation}
where $g$ is a coupling $\sim O(1~$eV), with both dipolar strain and
electric dipole contributions. The length $a_0$ is a typical lattice
distance; and we have suppressed an angular factor which has roughly
dipolar symmetry. We note that all of the principal scenarios for
the low-T behaviour of glasses assume (\ref{dipint}) to be true;
moreover they rely upon it in an essential way.

But is (\ref{dipint}) really correct? In this paper we shall argue
that in fact (\ref{dipint}) is incomplete, and that the correct form
contains extra terms of some importance. These include another term
falling off like $1/R_{12}^3$, which however leads to a random field
acting on each tunneling rotator. There are also terms which are
weaker and which fall off faster, which would be less important
except that they act on 2-level systems that do not see the first
random field. The net result of this is that we derive two effective
Hamiltonians for the system, one which is valid at higher
temperatures around $T_G$ and the other at much lower $T$,
apparently around the temperature $T_Q$ which defines the crossover
to the universal properties.

%%%%%%%%%%%%%%%%%%%%%%%%%%%%%%%%%%%%%%%%%%%%%%%%%%%%%%%%%%%%%%%%%%%%%%%%%%%%%%%%%%%%%%%%%%%%%%%%%%
%%%%%%%%%%%%%%%%%%%%%%%%%%%%%%%%%%%%%%%%%%%%%%%%%%%%%%%%%%%%%%%%%%%%%%%%%%%%%%%%%%%%%%%%%%%%%%%%%%
\section{Defect-Phonon Interactions}
 \label{sec:bareH}
%%%%%%%%%%%%%%%%%%%%%%%%%%%%%%%%%%%%%%%%%%%%%%%%%%%%%%%%%%%%%%%%%%%%%%%%%%%%%%%%%%%%%%%%%%%%%%%%%%
%%%%%%%%%%%%%%%%%%%%%%%%%%%%%%%%%%%%%%%%%%%%%%%%%%%%%%%%%%%%%%%%%%%%%%%%%%%%%%%%%%%%%%%%%%%%%%%%%%

In what follows we will be dealing with neutral glasses, i.e., we
ignore metallic and superconducting glasses. This of course still
includes the overwhelming majority of materials on earth, from rocks
and minerals to a galaxy of insulating compounds (based mostly on
transition metals), along with a huge variety of natural and
artificial organic systems (including polymers).

In spite of this variety, and regardless of whether one is dealing
with a strongly disordered amorphous systems or very weakly
disordered systems like substitutional electrolyte glasses, the two
interactions of main interest are those involving strain fields and
phonons, and those involving the interaction of electric fields with
local charge distributions.

%%%%%%%%%%%%%%%%%%%%%%%%%%%%%%%%%%%%%%%%%%%%%%%%%%%%%%%%%%%%%%%%%%%%%%%%%%%%%%%%%%%%%%%%%%%%%%%%%
\subsection{Simple model for Defects}
 \label{sec:def}
%%%%%%%%%%%%%%%%%%%%%%%%%%%%%%%%%%%%%%%%%%%%%%%%%%%%%%%%%%%%%%%%%%%%%%%%%%%%%%%%%%%%%%%%%%%%%%%%%

Our tactic in this paper will be to start with a toy model which
describes a class of very simple systems, and then argue that the
important features of this model can be generalised to a much wider
variety of glassy systems.

\vspace{2mm}

\begin{figure}[ht!]
\subfigure[]{\includegraphics[width =
0.3\columnwidth]{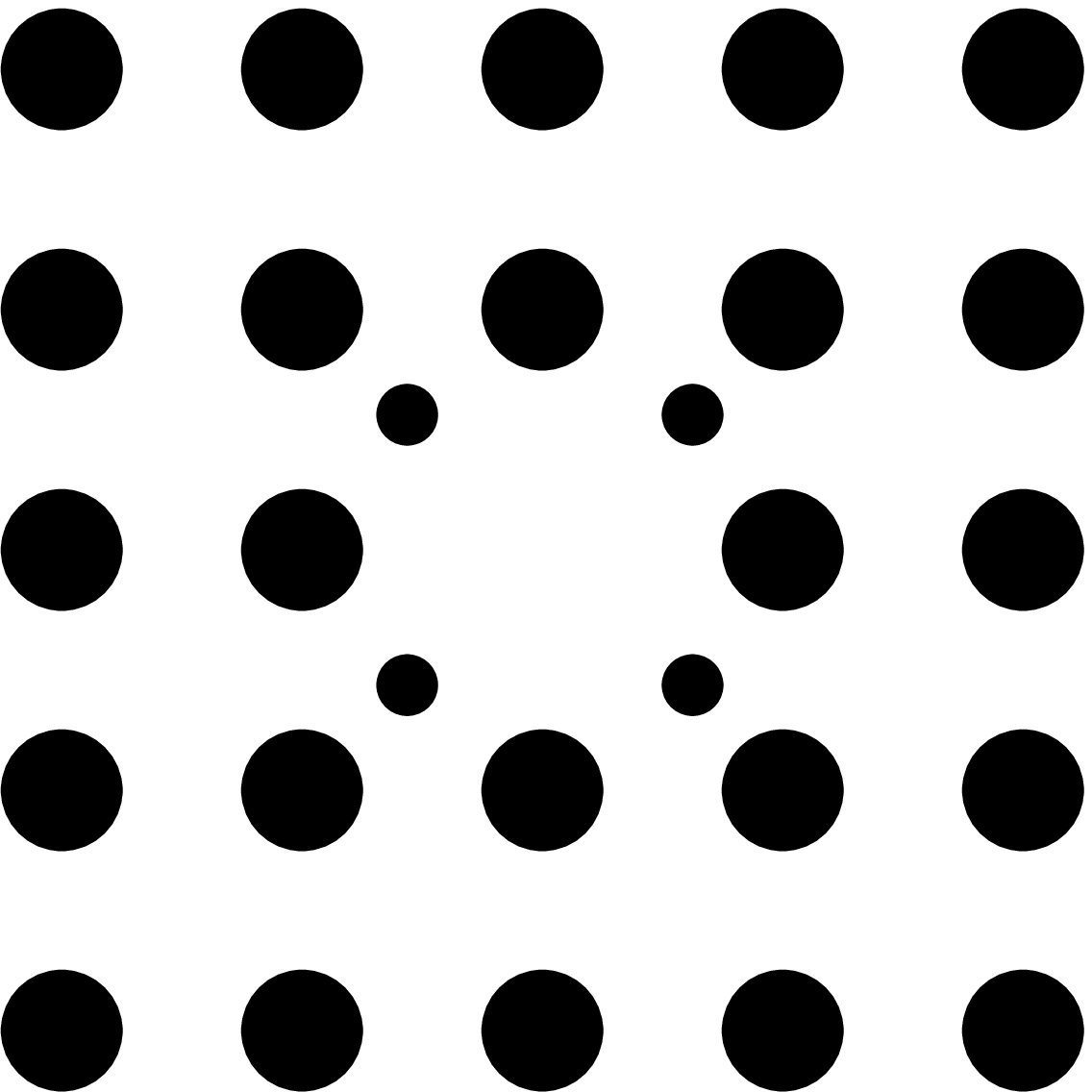}} \hspace{25mm}
\subfigure[]{\includegraphics[width =
0.3\columnwidth]{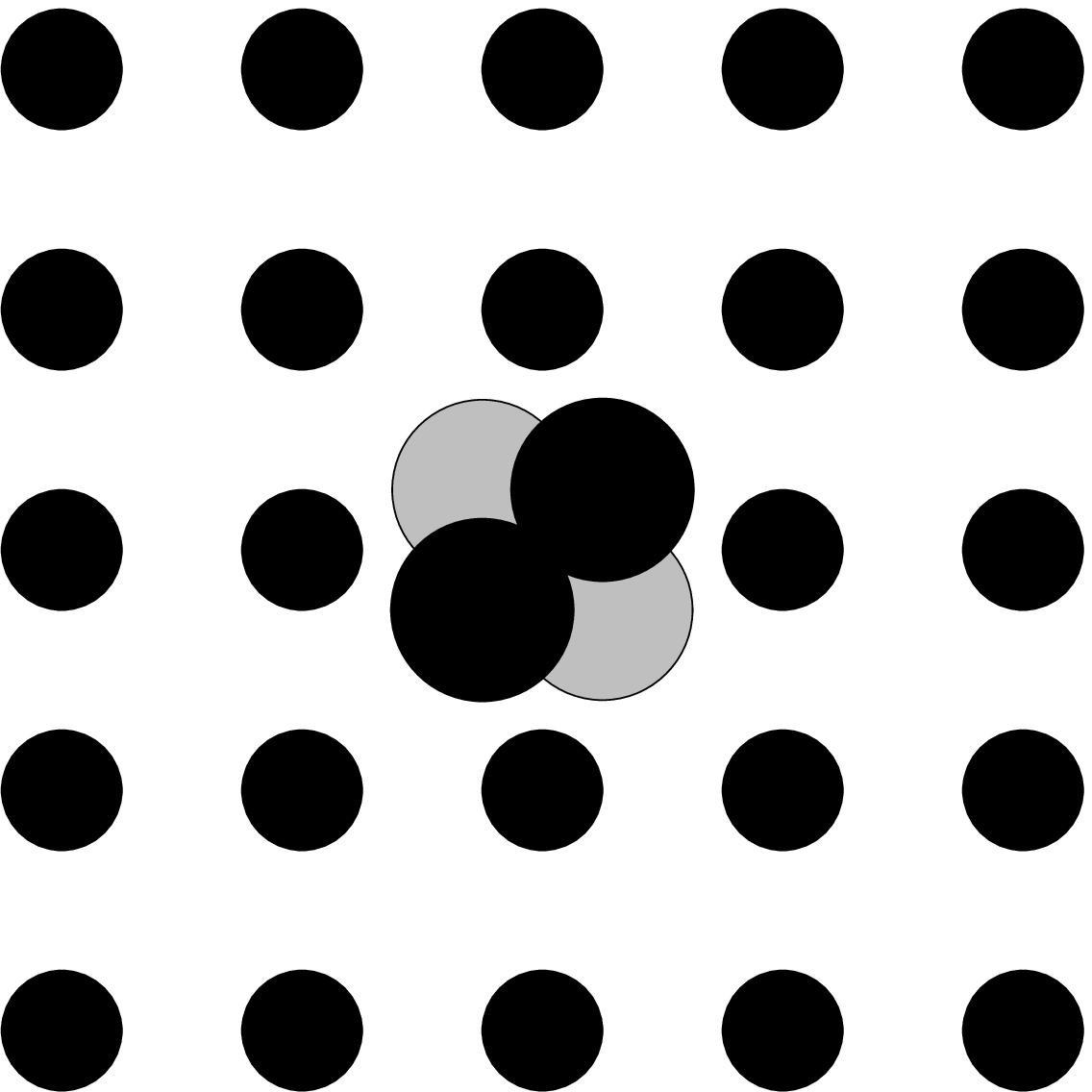}}
 \caption{(a) Off-center impurity with $4$ equivalent sites and
 (b) orientational impurity in the dumbbell approximation,
 with $2$ equivalent states. To first order in the displacement,
 the two opposite sites in (a) are also equivalent, leading to an
 effective $2$-state impurity. }
    \label{fig1}
\end{figure}

\vspace{6mm}

Consider the situation depicted in Fig. \ref{fig1}, in which we
reduce the system to a 2-d plane, and look at defects in this plane.
The underlying local symmetry of the system is assumed to be of
square plaquettes, and the defects can either be substitutional
defects, able to occupy one of 4 states in the plaquette, or
orientational defects, able to rotate between 4 orientational
states. Under certain circumstances, to be discussed below, we can
make the 'dumbbell approximation, in which defects rotated by
$180^o$ are considered to be indistinguishable - we then treat these
2 states as identical, and the 4-state system reduces to a 2-state
system.

\begin{figure}
\begin{center}
\includegraphics[width = 0.4\columnwidth]{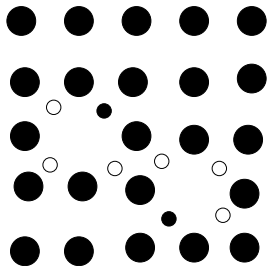}
 \caption{ Two nearby impurities, with distorted plaquettes.
 Filled small circles denoted occupied impurity sites. }
    \label{fig2}
    \end{center}
\end{figure}

Now if the concentration of these defects is low we can assume that
they do not disturb the underlying lattice symmetry, and
interactions between 2 defects, even if they are distantly
separated, will be between 2 plaquettes which are oriented along the
same axes. More generally, when the defect concentration is much
higher, one may have the situation shown in Fig. \ref{fig2}, where
the 2 plaquettes are slightly distorted, and also rotated with
respect to each other.

It might be objected that the situation depicted in Fig. \ref{fig2}
is not very realistic, in that a high defect concentration would so
severely disrupt the square symmetry that the plaquettes themselves
would not only be rotated, but that their shapes would be severely
distorted, so that no clear local lattice structure could be
defined. Actually this is not the case - quite surprisingly, the
situation in even rather strongly amorphous glasses does not conform
to the common caricature in which they look like frozen liquids.
Instead, at short length scales the underlying lattice structure is
still quite recognizable, and the system more resembles a 'frozen
liquid' of very small polycrystals\cite{Gas79} (actually the
instantaneous state of quite a few liquids also looks like this!).

\vspace{2mm}

\vspace{6mm}

A good tutorial example of systems like our toy model is provided by
substitutional electrolyte glasses, where the defects can be
characterized very precisely. Canonical examples are ${\rm
KCl_{1-x}Li_x}$, ${\rm KCl_{1-x}(OH)_x}$, ${\rm KBr_{1-x}(CN)_x}$,
and so on\cite{LM94,VG90}. Many experiments in these systems have
been done in the very dilute regime, with concentrations in the
range $10^{-6} < $ x $ < 5 \times 10^{-5}$, where one can ignore the
interactions between the widely spaced
impurities\cite{HPP68,Wal67,EGH+96}. However there are also many
studies of concentrations up to x $ \sim 0.1$ or even higher. When
$x > 2 \times 10^{-4}$, interactions between the impurities clearly
dominate the physics in most of these
systems\cite{VG90,EGH+96,WHE99,BS68}. This is shown in, e.g., the
saturation in the dielectric response for higher
concentrations\cite{EGH+96}.

In real 3-dimensional systems like ${\rm (KCl)_{1-x}Li_x}$, ${\rm
(KCl)_{1-x}(OH)_x}$, or ${\rm (KBR)_{1-x}(CN)_x}$, one has either
off center point-like impurity states (as in ${\rm
(KCl)_{1-x}Li_x}$, with 8 available positions for the ${\rm Li}$
impurity inside a given ${\rm KCl}$ lattice 'cage'); or else
orientational (as in ${\rm (KBR)_{1-x}(CN)_x}$, where the ${\rm CN}$
impurity can lie in one of the 8 directions along (1,1,1) and its
equivalents). Previous treatments of the microscopic interactions in
these systems have taken one of two routes. Michel and
collaborators\cite{Mic86,LM94} have used a detailed microscopic
description of the orientational and translational degrees of
freedom (and the coupling between the two of these) with the goal of
describing systems where orientational impurities like ${\rm CN}$
can rotate inside their host 'cages'. Sethna and
collaborators\cite{GRS90,SC85,Set86} have used a somewhat more
general phenomenological description in terms of TLS impurities.

The great advantage of beginning with these simple systems is that
one can set up a well-controlled theory beginning with the case of
dilute defect concentration $x$. There are good arguments, based on
the experiments on electrolyte
glasses\cite{VG90,HPP68,Wal67,EGH+96,WHE99,BS68} and other amorphous
systems\cite{PLT02,LNRO03,LO03,NFHE04,WEL+95,BNOK98,HR86}, that the
theory must, when $x$ is larger, flow towards the strong coupling
regime (and perhaps to some universal quantum regime). But it makes
no sense trying to explore the strong-coupling limit until the form
of the interactions has been established in weak coupling. The key
question of interest here (viz., what are these interactions?) can
only rigorously be addressed by starting from a system where x$ \ll
1$. Later we will argue that our main results will survive well into
the universal regime, and so are of much more general applicability.

Returning now to Fig. \ref{fig1}, we divide the defects into two
groups, viz. (i) those in which there is an inversion symmetry
relating pairs of states, where inversion is made with respect to
the relevant lattice position; and (ii) those where there is no such
symmetry. In the square plaquette system we see that both simple
impurities like ${\rm Li}$ and orientational impurities like ${\rm
CN}$ or ${\rm OH}$ can be in one 4 different states, related by
$90^o$ rotations of the plaquette, and there is no inversion
symmetry. However in some cases, one can treat the states related by
$180^o$ as physically equivalent (either because they really are
equivalent, or because at the energy scale of interest the
difference is unimportant), and then we can assume inversion
symmetry. In this case the state space on the plaquette is
2-dimensional, with states oriented along one or other of the
lattice diagonals.

To describe the system, the simplest representation is just a
4-state one in which the system can hop between any of the 4
plaquette sites. However we will use a slightly different one, where
we begin by defining a set of operators $\{ \hat{\tau}_i \}$ such
that $\hat{\tau}_i^x$ flips the state on the plaquette through
$180^o$, and another set of operators $\{ \hat{S}_i \}$ such that
$\hat{S}_i^x$ rotates the state on the plaquette through $90^o$ in
either direction. This representation is used because we will see
that the impurity-lattice interaction turns out to depend crucially
on whether inversion symmetry is obeyed, and so this way of setting
up the description allows us to distinguish between operations which
are or are not invariant under inversion symmetry.

In what follows we will first be considering the effect of phonons
on the rotational tunneling degrees of freedom. In general, before
we take interactions with phonons into account, these will be
described by a Hamiltonian
\begin{equation}
H_{\rm def}^{(S)} = \sum_{j} E_j \hat{S}_j^z  - D_j \hat{S}_j^x
 \label{HdeftS}
\end{equation}
where $D_j$ is a tunneling matrix element and $E_j$ is any stray
field acting on the rotational defect variable $\hat{S}_j$. At lower
$T$, once the rotational degrees of freedom are frozen out, only the
inversion tunneling processes are left, and these are described by a
bare Hamiltonian
\begin{equation}
H_{\rm def}^{(\tau)} = \sum_{j} \epsilon_j \hat{\tau}_j^z  -
\Delta_j \hat{\tau}_j^x
 \label{Hdefttau}
\end{equation}
where again an "external" effective longitudinal field $\epsilon_j$
acting on the $j$-th dipole is allowed.

Often, in the study of the possible phases of the system, the
high-$T$ tunneling is dropped since it is typically too small to
influence the nature of the phase, nor its stability to effective
random fields\cite{TH95}. We will keep it simply because we wish to
trace how the effective Hamiltonian evolves as we lower the energy
scale.

%%%%%%%%%%%%%%%%%%%%%%%%%%%%%%%%%%%%%%%%%%%%%%%%%%%%%%%%%%%%%%%%%%%%%%%%%%%%%%%%%%%%%%%%%%%%%%%%%
\subsection{Coupled Defect-Phonon System}
 \label{sec:phon}
%%%%%%%%%%%%%%%%%%%%%%%%%%%%%%%%%%%%%%%%%%%%%%%%%%%%%%%%%%%%%%%%%%%%%%%%%%%%%%%%%%%%%%%%%%%%%%%%%

Consider now the coupling to the phonon modes in the system. The
Hamiltonian for this system is given by:
\begin{equation}
H=H_{\rm def} + H_{\rm ph} + H_{\rm int}
 \label{Heff}
\end{equation}
where $H_{\rm def}$ is the bare defect term just discussed, and the
phonon system is described by
\begin{equation}
H_{\rm ph} = \sum_{q,\mu} \left( \frac{|P_{q,\mu|^2}}{2M} + M
\omega_{q,\mu}^2 \frac{|X_{q,\mu|^2}}{2} \right) .
 \label{phonon}
\end{equation}
Here $P$ and $X$ represent momentum and displacement operators for
phonon modes of wave vector $q$ and branch $\mu$, and $M$ is the
mass of the elementary cell of the medium.

We split the defect-phonon interaction $H_{\rm int}$ into two terms,
writing
\begin{equation}
H_{\rm int} = H^{\rm int}_1 + H^{\rm int}_2
 \label{Hint}
\end{equation}
where the lowest order term is linear in the phonon displacement
field:
\begin{equation}
H^{\rm int}_1 = -\sum_{\alpha,\beta} \sum_{j} \left( \eta ({\bf
r}_j) \delta^{\alpha \beta} \frac{\partial X_{j \alpha}}{\partial
x_{j \alpha}} \;+\; \gamma_{\alpha \beta}({\bf r}_j) \frac{\partial
X_{j \alpha}}{\partial x_{j \beta}} S_j^z  \right)
 \label{H1int}
\end{equation}
and the next term is a non-linear coupling of the defects to the
gradient of this field, of form
\begin{equation}
H^{\rm int}_2 = -\sum_{\alpha,\beta,\delta} \sum_{j} \zeta_{\alpha
\beta \delta}({\bf r}_j) \frac{\partial^2 X_{j \alpha}}{\partial
x_{j \beta}
\partial x_{j \delta}} \tau_j^z .
 \label{H2int}
\end{equation}
The linear coupling term $H^{\rm int}_1$ contains first a 'volume
coupling' with coefficient $\eta$, which is independent of the
defect position or orientation, arising because the defect has a
different volume from the host and this locally strains it. Then
there is the usual interaction $\gamma_{\alpha \beta}({\bf r}_j)$
between the defect and the strain field in the "TLS" or "dumbbell"
approximation, which changes sign with $S_j^z$, since the phonon
displacement fields are sensitive to the defect orientation. The
size of these interactions is not easy to calculate - however
$\gamma_{\alpha \beta}({\bf r}_j)$ can be measured, and is on a
typical scale $\gamma \sim 1~$ eV. Estimates of $\eta$ give similar
numbers but the ratio $\eta/\gamma$ must certainly vary from one
system to another (calculations are difficult because the volume
change and charge redistributions caused by the defect will interact
with each other).

We notice that when inversion symmetry is preserved, there is no
linear interaction between the $\hat{\tau}_j$ and the phonon field.
The 2nd term $H^{\rm int}_2$ arises when we relax the assumption of
inversion symmetry, so states rotated by $180^o$ are distinguishable
(either because we relax the "dumbbell" approximation, and consider
the dipole moment of the impurity or molecule, or because we
consider tunneling point impurities). In this case the effective
'dipole' represented by the difference between these states can
interact with the gradient of the phonon displacement field, in the
form given in $H^{\rm int}_2$. However this interaction is much
weaker, and has rarely been considered before. Certainly we do not
expect it to affect the physics near the glass freezing temperature.
However we will see that this interaction is important at lower
temperatures, where other degrees of freedom freeze out. At the
present time we cannot give more than a rough estimate for the size
of the coefficient $\zeta^{\alpha \beta \delta}$ of this
interaction; noting that its dimensions are ${\cal D}[\zeta^{\alpha
\beta \delta}] = EL$ (ie., energy times length), and that the
characteristic energy scale of defect interaction energies is $U_o
\sim 1~$ eV, and the characteristic length scale $a_d$ of defect
dynamics is roughly $1~{\AA}$, we would expect that $\vert
\zeta^{\alpha \beta \delta} \vert \sim 1~eV \times {\AA}$. To get an
energy from this we need to divide by the typical distances between
impurities, given by $l_d \sim a_o x^{-1/3}$, where $a_o$ is a
lattice length. Thus we expect that the energy scale associated with
this coupling is less than $\gamma$ and $\eta$ by a factor $a_d
x^{1/3}/a_o$. From now on we will assume that this energy scale
associated with $\zeta$ is considerably smaller than $\gamma,\eta$.

%%%%%%%%%%%%%%%%%%%%%%%%%%%%%%%%%%%%%%%%%%%%%%%%%%%%%%%%%%%%%%%%%%%%%%%%%%%%%%%%%%%%%%%%%%
%%%%%%%%%%%%%%%%%%%%%%%%%%%%%%%%%%%%%%%%%%%%%%%%%%%%%%%%%%%%%%%%%%%%%%%%%%%%%%%%%%%%%%%%%%
\section{Inter-defect Effective Hamiltonian: High Energies}
 \label{sec:dumb}
%%%%%%%%%%%%%%%%%%%%%%%%%%%%%%%%%%%%%%%%%%%%%%%%%%%%%%%%%%%%%%%%%%%%%%%%%%%%%%%%%%%%%%%%%%
%%%%%%%%%%%%%%%%%%%%%%%%%%%%%%%%%%%%%%%%%%%%%%%%%%%%%%%%%%%%%%%%%%%%%%%%%%%%%%%%%%%%%%%%%%

Let us begin by assuming that we are at a sufficiently high energy
scale that we can neglect the weaker interaction $H^{int}_2$; we are
then concerned with a set of phonons interacting with the $\{
\hat{S}_j \}$ TLS variables, ignoring the $\{ \hat{\tau}_j \}$
variables. We wish, in this approximation, to calculate the
effective Hamiltonian of the system, initially just up to 2nd-order
in the defect-phonon coupling. This is an old problem, but we shall
see that even here there are new things to be discovered. Let's look
first at the simple problem of 2 interacting defects in this
dumbbell approximation. We split the interaction term $H^{\rm
int}_1$ into two, to understand the effect of the orientation and
volume parts separately, and consider 2 defects at positions ${\bf
r}_1$ and ${\bf r}_2$. Then the system has bare Hamiltonian:
\begin{eqnarray}
H^{\rm int}_1 &=& V_{\gamma} + V_{\eta} \nonumber \\
&=& -\sum_{\alpha,\beta} \left(\gamma_1^{\alpha \beta}
\frac{\partial X_{1 \alpha}}{\partial x_{1 \beta}} \hat{S}_1^z +
\gamma_2^{\alpha \beta} \frac{\partial X_{2 \alpha}}{\partial x_{2
\beta}} \hat{S}_2^z \right) \nonumber \\ & & \;-\; \sum_{\alpha}
\left( \eta_1 \frac{\partial X_{1 \alpha}}{\partial x_{1 \alpha}} +
\eta_2 \frac{\partial X_{2 \alpha}}{\partial x_{2 \alpha}} \right) .
 \label{Hint2}
\end{eqnarray}
If we now integrate out the phonons we will generate, at lowest
order in $H^{\rm int}_1$, terms of Ising form (proportional to
$\gamma_1 \gamma_2 \hat{S}_1^z \hat{S}_2^z$), cross terms giving a
local field, proportional to $\gamma \eta (\hat{S}_1^z +
\hat{S}_2^z)$, plus an energy shift $\sim \eta_1 \eta_2$. The Ising
form has been known for a long time. The second term, when summed
over all spins apart from one given spin, simply leads to a random
field at the site of that spin - this term is not usually
considered. In this section we first sketch the derivation of the
Ising term, primarily to establish notation, and then derive the
random field term.

%%%%%%%%%%%%%%%%%%%%%%%%%%%%%%%%%%%%%%%%%%%%%%%%%%%%%%%%%%%%%%%%%%%%%%%%%%%%%%%%%%%%%%%%%%%
\subsection{Ising interaction term}
 \label{sec:Ising}
%%%%%%%%%%%%%%%%%%%%%%%%%%%%%%%%%%%%%%%%%%%%%%%%%%%%%%%%%%%%%%%%%%%%%%%%%%%%%%%%%%%%%%%%%%%

We define the Fourier transformation to momentum space as
\begin{equation}
X_{1 \alpha}(x) = \frac{1}{\sqrt{N}} \sum_{q,\mu} X_{q, \mu} {\bf
e}_{q, \mu, \alpha} e^{i q x} \label{Fourier}
\end{equation}
where ${\bf e}_{q, \mu, \alpha}$ is a phonon polarisation index.
Then we have
\begin{equation}
V_{\gamma} = - \frac{1}{\sqrt{N}} \sum_{\alpha,\beta} \gamma_{1
\alpha \beta} \sum_{q,\mu} X_{q, \mu} {\bf e}_{q, \mu, \alpha} i
q_{\beta} e^{i q x} S_1^z \;+\; (1 \leftrightarrow 2) .
\end{equation}

To find the interaction in 2nd-order perturbation theory we minimize
the potential energy, i.e. the sum of the second term in
Eq.(\ref{phonon}) plus the interaction term. Straightforward
calculation results in energy terms proportional to $(S_1^z)^2,
(S_2^z)^2$ and the Ising interaction term of interest, proportional
to $\hat{S}_1^z \hat{S}_2^z$. Let us define the notation ${\bf
R}_{12} = {\bf x}_1 - {\bf x}_2$, and use an acoustic approximation,
in which the longitudinal phonon frequency $\omega_{q l} = c_l q$,
the transverse phonon frequency is $\omega_{q \perp} = c_{\perp} q$.
We also have the identities:
\begin{eqnarray}
&\bfe_{q l \alpha} = q_{\alpha}/q, \;\;\;\;\;\;\;\;
 \bfe_{q_{\perp1}} \cdot q = \bfe_{q_{\perp 2}} \cdot q =
\bfe_{q_{\perp 1}} \cdot \bfe_{q_{\perp 2}} = 0
\nonumber \\
&\sum_{\mu=\perp_1, \perp_2} \bfe_{q \mu \alpha} \bfe_{q \mu \beta}
= \delta_{\alpha \beta} - q_{\alpha} q_{\beta}/q^2 .
 \label{polI}
\end{eqnarray}
Then the Ising interaction $V_{12}^{zz} = U_{12}^{zz}\hat{S}_1^z
\hat{S}_2^z$, where
\begin{eqnarray}
U_{12}^{zz} &=& - \sum_{q, \mu} \frac{1}{NM\omega_{q \mu}^2}
\sum_{\alpha \beta \gamma \delta} \bfe_{q \mu \alpha} \bfe_{q \mu
\gamma} q_{\beta} q_{\delta} \gamma_1^{\alpha \beta}
\gamma_2^{\gamma \delta} \nonumber \\ & & \times \cos{[q(x_1 -
x_2)]} .
 \label{Uzz}
\end{eqnarray}
Summing over polarization indices gives
\begin{eqnarray}
& & U_{12}^{zz} = - \frac{1}{NM} \sum_{\alpha \beta \gamma \delta}
\gamma_1^{\alpha \beta} \gamma_2^{\gamma \delta}
\left(\frac{1}{c_l^2} - \frac{1}{c_{\perp}^2} \right) \nonumber
\\ & & \times \sum_q \frac{q_{\alpha} q_{\beta} q_{\gamma} q_{\delta}}{q^4}
\cos{({\bf q} \cdot {\bf R}_{12})} \nonumber
\\
&-& \frac{1}{NM}\sum_{\alpha, \beta, \delta}
 {\gamma_1^{\alpha \beta} \gamma_2^{\alpha \delta} \over c_{\perp}^2}
\sum_q {q_{\beta} q_{\delta} \over q^2} \cos{({\bf q} \cdot {\bf
R}_{12})}
 \label{Uzz-sum}
\end{eqnarray}
which when the sum over momenta is performed, gives the real space
form
\begin{equation}
U_{12}^{zz} = {g_{12}^{zz}  \over R_{12}^3}
 \label{U-g}
\end{equation}
with the interaction

\begin{eqnarray}
%\hspace{-20mm}
& & g_{12}^{zz} ({\bf n}) = {-1 \over 4\pi \rho c_{\perp}^2}
\gamma_1^{\alpha \beta} \gamma_2^{\alpha \delta}(\delta_{\beta
\delta} - 3 n_{\beta} n_{\delta}) \nonumber \\ &-& {1 \over 4 \pi
\rho}
 \left( {1 \over c_l^2} - {1 \over c_{\perp}^2} \right)
\gamma_1^{\alpha \beta} \gamma_2^{\gamma \delta} ( - \left[
\delta_{\alpha \beta} \delta_{\gamma \delta} + \delta_{\alpha
\gamma} \delta_{\beta \delta} + \delta_{\alpha \delta} \delta_{\beta
\gamma} \right] \nonumber \\
%\hspace{-20mm}
&+& 3 \left[ \delta_{\alpha \beta} n_{\gamma} n_{\delta} +
\delta_{\alpha \gamma} n_{\beta} n_{\delta} + \delta_{\alpha \delta}
n_{\beta} n_{\gamma} + \delta_{\beta \gamma} n_{\alpha} n_{\delta}
\right] \nonumber \\ &+& 3 \left[\delta_{\beta \delta} n_{\alpha}
n_{\gamma} + \delta_{\gamma \delta} n_{\alpha} n_{\beta} \right] -
15 n_{\alpha} n_{\beta} n_{\gamma} n_{\delta}) \; \; \; \;
 \label{Uzz-real}
\end{eqnarray}

Here the reduced variable ${\bf n} = {\bf R}/\vert {\bf R} \vert$,
and we suppress the "$\{12\}$" subscript on ${\bf R}_{12}$ and ${\bf
n}_{12}$ to keep things uncluttered. This interaction has a rather
complicated angular dependence, coming both from the anisotropy of
the medium, and from the that of the strain interaction
$\gamma_{\alpha \beta}({\bf r})$. If we assume a completely
isotropic medium with degenerate longitudinal and transverse
phonons, and also make the simplification of anisotropic
$\gamma_{\alpha \beta} ({\bf r})$, so that $\gamma_{\alpha
\beta}({\bf r}) \rightarrow \gamma_o \delta_{\alpha z} \delta_{\beta
z}$, we get the strictly dipolar form
\begin{equation}
g^{zz} ({\bf n}) \;=\; \left({\gamma_o^2 \over 4 \pi \rho
c_o^2}\right) \; [ 3 \cos^2 \theta({\bf n}) - 1]
 \label{g_o}
\end{equation}
where $\theta({\bf n})$ is the angle between the unit radius vector
${\bf n}$ and the $\hat{z}$-axis; the characteristic coupling energy
$g_o = (\gamma_o^2/ \pi \rho c_o^2)$ is now evident.

As noted before, these results are well-known, and were first
derived\cite{JL75,BH77} in the 1970's. The essential result is that
one has derived an effective Ising interaction; since the sites of
the defects are random, the tunneling terms make this system behave
as a quantum Ising model with random interactions, and Hamiltonian
\begin{equation}
H = \sum_i D_i \hat{S}_i^x \;+\; \sum_{ij} U_{ij}^{zz} \hat{S}_i^z
\hat{S}_j^z
 \label{QIsing}
\end{equation}
where the tunneling amplitudes $\{ D_i \}$ are typically much
smaller than the nearest-neighbour Ising interactions. However it
turns out that this Ising interaction is not the only term that is
important.

%%%%%%%%%%%%%%%%%%%%%%%%%%%%%%%%%%%%%%%%%%%%%%%%%%%%%%%%%%%%%%%%%%%%%%%%%%%%%%%%%%%%%%%%%%
\subsection{Random Field term}
 \label{sec:RFT}
%%%%%%%%%%%%%%%%%%%%%%%%%%%%%%%%%%%%%%%%%%%%%%%%%%%%%%%%%%%%%%%%%%%%%%%%%%%%%%%%%%%%%%%%%%

We now include the cross-terms $\propto \gamma \eta $, coming from
2nd-order perturbation theory in the interaction $H^{\rm int}_1$ in
Eq.(\ref{Hint2}). Using similar manouevres as in the calculation
above, one now obtains another term in the effective interaction of
form $V_{12}^{z0} = U_{12}^{z0}(\hat{S}_1^z + \hat{S}_2^z)$, where
\begin{equation}
U_{12}^{z0}({\bf R}_{12}) = -\frac{1}{NM} \sum_{\alpha \beta}
\frac{\eta \gamma_{\alpha \beta}}{c_l^2} \sum_q \frac{q_{\alpha}
q_{\beta}}{q^2} \cos{({\bf q} \cdot {\bf R}_{12})} .
\end{equation}
We call this a random field term because if we take a given spin in
the system, say $\hat{S}_i$, and then sum the interaction
$U_{ij}^{z0}({\bf R}_{ij})$ between $\hat{S}_i$ and the volume terms
coming from all the other defect sites $\{ j \}$, we get a field
$h_i^z$ acting on $\hat{S}_i^z$ at site $i$ which varies from site
to site in a random way, because of the random positions and
orientations of the arguments ${\bf R}_{ij}$. Note, incidentally,
that this effective random field interaction has contribution only
from the longitudinal phonons.

This interaction is important since, as we show elsewhere, it
actually destroys the bulk glass transition\cite{SS06}. Let us now
explicitly derive its form. To do this we first evaluate the tensor
\begin{equation}
\tilde{I}_{\alpha \beta} \;=\; \sum_q \frac{q_{\alpha}
q_{\beta}}{q^2} \cos{(q \cdot R)} .
\end{equation}
Changing the sum to an integral we get:
\begin{equation}
\tilde{I}_{\alpha \beta} \;=\; \frac{V}{(2 \pi R)^3} \int d^3q
\frac{q_{\alpha} q_{\beta}}{q^2} \cos{(q \cdot \bfn)} \;\equiv \;
\frac{V}{(2 \pi R)^3} I_{\alpha \beta}
\end{equation}
where again $\bfn \equiv \bfR/|\bfR|$. From symmetry we have
$I_{\alpha \beta} = a \delta_{\alpha \beta} + b \bfn_{\alpha}
\bfn_{\beta}$. Consider first $\sum_{\alpha} I_{\alpha \alpha}$. On
one hand this sum can be written as $\int d^3q \cos{(q \cdot \bfn)}
= 0$. On the other hand the sum equals $3a + b$, leading to the
identity $3a+b=0$.

Similarly, the scalar
\begin{equation}
\sum_{\alpha \beta} I_{\alpha \beta} \bfn_{\alpha} \bfn_{\beta} =
\int d^3q \frac{q_{\alpha} q_{\beta} \bfn_{\alpha}
\bfn_{\beta}}{q^2} \cos{(q \cdot \bfn)} .
\end{equation}
In calculating the right side we can choose $\bfn$ in any direction,
say in the z direction. This leads to the expression
\begin{equation}
2 \pi \int dq_r q_r \int dq_z \frac{q_z^2}{q_r^2 + q_z^2} \cos{q_z}
.
\end{equation}
and then using
\begin{equation}
\frac{q_z^2}{q_r^2 + q_z^2} = 1 - \frac{q_r^2}{q_r^2 + q_z^2}
\end{equation}
and straightforward integration we find that the integral equals $-4
\pi^2$. on the other hand, $\sum_{\alpha \beta} I_{\alpha \beta}
\bfn_{\alpha} \bfn_{\beta} = a + b$, leading to the identity $a+b=-4
\pi^2$, and together with the above identity ($3a+b=0$) to the
result that $a=2 \pi^2, b=-6 \pi^2$, and
\begin{equation}
I_{\alpha \beta} = -2 \pi^2 (3 \bfn_{\alpha} \bfn_{\beta} -
\delta_{\alpha \beta}) .
\end{equation}

Thus, one finally obtains
\begin{equation}
U_{12}^{z0} = \frac{1}{4 \pi \rho c_l^2 R^3} \sum_{\alpha \beta}
\eta \gamma^{\alpha \beta} \; (3 \bfn_{\alpha} \bfn_{\beta} -
\delta_{\alpha \beta}) .
\end{equation}
This result shows a much less complicated angular dependence than
the Ising interaction (\ref{Uzz-real}); in general we see that they
depend differently on angle, simply because the Ising $g^{zz}({\bf
r})$ is essentially a dipole-dipole interaction whereas this mixed
term is a dipole-monopole interaction.

If we make the isotropic assumption that $\gamma_{\alpha \beta}({\bf
r}) \rightarrow \gamma_o$, then $U_{12}^{z0}  \; \rightarrow \; 0$.
On the other hand if $\gamma_{\alpha \beta} = \gamma_{zz}$, then we
get a dipolar interaction:
\begin{equation}
U_{12}^{z0}  \; \rightarrow \; {g^{z0} ({\bf n}) \over R^3}
 \label{U-z0o}
\end{equation}
where
\begin{equation}
 g^{z0}({\bf n}) \; = \;
 \left( {\gamma_{zz} \eta \over 4 \pi \rho c_l^2} \right)
\; [ 3 \cos^2 \theta({\bf n}) - 1]
 \label{Uh-iso}
\end{equation}
where now the interaction energy is $( \gamma_{zz} \eta/ 4 \pi \rho
c_l^2)$. In any case, both this mixed term and the Ising term end up
having the same $1/R_{ij}^3$ spatial form, but their characteristic
energies are different.

Thus, as a result of the added volume term in the impurity-lattice
interaction, the effective Hamiltonian in the "dumbbell"
approximation is not the quantum Ising model (\ref{QIsing}), but the
quantum random field Ising model, with Hamiltonian
\begin{equation}
H_{RF} ( \{ \hat{S}_j \}) \;=\; \sum_{ij} U_{ij}^{zz} \hat{S}_i^z
\hat{S}_j^z + \sum_i (D_j \hat{S}_i^x \;+\; B_i \hat{S}_i^z)
 \label{RFIsing}
\end{equation}
where $B_i = \sum_j U_{ij}^{z0}$. This random field has mean zero,
as by symmetry there is no preferred direction. Its typical value is
given by $B_0 \approx \eta \gamma {\rm x}/(4 \pi \rho c_l^2)$, where
x is the concentration, since it is dictated by the typical distance
between nearest impurities. Thus, the typical size of both
interactions is rather similar; we expect that $\vert B_i \vert \sim
\vert \sum_j U_{ij}^{zz} \vert (\eta/\gamma)$. In typical glassy
systems this means that they are both $\sim {\rm x} U_o$, where $U_o
\sim 1~$ eV. This means that unless ${\rm x} < 10^{-4}$, the typical
size of these random fields $\vert B_i \vert \gg D_o$, where $D_o$
is a typical tunneling amplitude. Note, that the standard deviation
of the distribution of the random fields, $\bar{B} \approx \eta
\gamma \sqrt{\rm x}/(4 \pi \rho c_l^2) \gg B_0$, as it is dominated
by the rare events of pairs of impurities occupying nearest neighbor
lattice points (see Ref.~\cite{Sch08}, noting the trivial relations
between high and low impurity densities).

The Random field Quantum Ising Hamiltonian (\ref{RFIsing}) has very
different properties from the simple quantum Ising system - apart
from anything else, the random field can actually destroy the glass
transition\cite{SS06}. We do not go into these questions here, but
note instead that the most important effect of the Ising interaction
term, for all but very dilute glasses, is to freeze the tunneling of
the $\hat{S}_j$ variables, except for a very small fraction $\sim
D_o/xU_o$ of systems that happen to be on resonance.

However this is not the end of the story at all. This is because, as
noted above, the $\hat{\tau}_j$ variables do not have a linear
coupling to the phonons, and so to linear order they experience
neither impurity-impurity interactions nor a random field, and thus
they are still free variables. This is why we now have to go to the
higher coupling terms.

%%%%%%%%%%%%%%%%%%%%%%%%%%%%%%%%%%%%%%%%%%%%%%%%%%%%%%%%%%%%%%%%%%%%%%%%%%%%%%%%%%%%%%%%%%%%%%%%%%%%
%%%%%%%%%%%%%%%%%%%%%%%%%%%%%%%%%%%%%%%%%%%%%%%%%%%%%%%%%%%%%%%%%%%%%%%%%%%%%%%%%%%%%%%%%%%%%%%%%%%%
\section{Inter-Defect Interactions at low Energy}
 \label{sec:dumb+}
%%%%%%%%%%%%%%%%%%%%%%%%%%%%%%%%%%%%%%%%%%%%%%%%%%%%%%%%%%%%%%%%%%%%%%%%%%%%%%%%%%%%%%%%%%%%%%%%%%%%
%%%%%%%%%%%%%%%%%%%%%%%%%%%%%%%%%%%%%%%%%%%%%%%%%%%%%%%%%%%%%%%%%%%%%%%%%%%%%%%%%%%%%%%%%%%%%%%%%%%%

Let us now go to an energy scale very much less than the putative
glass freezing temperature. Now we start from a Hamiltonian given by
\begin{eqnarray}
\hspace{-3mm} H_{\rm int} &=& H_{RF}(\{ {\bf S}_j \}) \;+\; \sum_j
\Delta_j \hat{\tau}_j^x - \sum_{\alpha,\beta,\delta} \sum_{j}
 \zeta_j^{\alpha \beta \delta} \frac{\partial^2
X_{j \alpha}}{\partial x_{j \beta} \partial x_{j \delta}} \tau_j^z
\nonumber \\ &-& \sum_{j,\alpha,\beta} \gamma_j^{\alpha \beta}
\frac{\partial X_{j \alpha}}{\partial x_{j \beta}} S_j^z \;-\;
\eta_j \sum_{\alpha} \frac{\partial X_{j \alpha}}{\partial x_{j
\alpha}}
 \label{fullinteractionH}
\end{eqnarray}
where the tunneling amplitude $\Delta_j$ describes the $180^o$ flip
of defects. The first term in this effective Hamiltonian is just the
random field Quantum Ising model derived in (\ref{RFIsing}) above.

Now let us write an approximate version of (\ref{fullinteractionH}),
which takes account of the fact that the set of spins $\{ {\bf S}_j
\}$ have almost entirely frozen into some random configuration, with
expectation values $\{ \langle {\bf S}_j \rangle \}$, because of the
strong Ising interaction between them (but without long-range glassy
order). Now in this approximation we can simply ignore the dynamics
of the $\{ {\bf S}_j \}$ entirely, and replace
(\ref{fullinteractionH}) with another Hamiltonian, which is
approximately valid when $T \ll T_G$, given by:
\begin{eqnarray}
H_{\rm int} &=&  \sum_j \Delta_j \hat{\tau}_j^x -
\sum_{\alpha,\beta,\delta} \sum_{j}
 \zeta_j^{\alpha \beta \delta} \frac{\partial^2
X_{j \alpha}}{\partial x_{j \beta} \partial x_{j \delta}} \tau_j^z
\nonumber \\ &-& \sum_{j,\alpha,\beta} \gamma_j^{\alpha \beta}
\frac{\partial X_{j \alpha}}{\partial x_{j \beta}} \langle S_j^z
\rangle \;-\; \eta_j \sum_{\alpha} \frac{\partial X_{j
\alpha}}{\partial x_{j \alpha}} .
 \label{truncint}
\end{eqnarray}
This Hamiltonian is only valid to the extent that we can ignore any
tunneling of resonant $\{ {\bf S}_j \}$ variables.

Consider now the interaction terms in Eq.(\ref{truncint}). If we now
integrate over the phonons, one expects the interaction between
$\hat{\tau}_j$ and the phonons [the 3rd term in
(\ref{fullinteractionH})] to give an interaction $\sim \zeta_i
\zeta_j \tau_i^z \tau_j^z$. But we notice that there will also be
two cross-terms between the two couplings to the phonons, giving an
interaction $\sim \gamma_i \zeta_j \tau_i^z \langle S_j^z \rangle$
and an interaction $\sim \eta_i \zeta_j \tau_i^z$. When summed over
the sites $\{ j \}$, both of these terms must give random fields
acting on the $\hat{\tau}_j$ variable. The first of these random
fields comes from the frozen $S_j$ degrees of freedom, behaving as a
quenched impurity distribution coupling to the $\tau^z$ degrees of
freedom. The second just comes from summing over all the scalar
volume distortions from these frozen impurities.

%%%%%%%%%%%%%%%%%%%%%%%%%%%%%%%%%%%%%%%%%%%%%%%%%%%%%%%%%%%%%%%%%%%%%%%%%%%%%%%%%%%%%%%%%%%%
\subsection{Interactions involving the  $\{ \hat{\tau}_j
\}$ variables}
 \label{sec:intSt}
%%%%%%%%%%%%%%%%%%%%%%%%%%%%%%%%%%%%%%%%%%%%%%%%%%%%%%%%%%%%%%%%%%%%%%%%%%%%%%%%%%%%%%%%%%%%

Since $\gamma, \eta \gg \zeta$, it follows that the two random field
terms acting on the $\{ \tau_j^z \}$ interaction will be much
stronger than the Ising interactions between them (quite different
from what occurs for the $\{ \hat{S}_j \}$). We therefore deal with
the two random field terms first, and then look at the Ising
interaction between the $\{ \tau_j^z \}$.

%%%%%%%%%%%%%%%%%%%%%%%%%%%%%%%%%%%%%%%%%%%%%%%%%%%%%%%
\subsubsection{The $S^z_i \tau^z_j$ interaction}
 \label{sec:Stau}
%%%%%%%%%%%%%%%%%%%%%%%%%%%%%%%%%%%%%%%%%%%%%%%%%%%%%%%

The $S^z_i \tau^z_j$ interaction is not only typically the largest
(although $\gamma$ and $\eta$ are of the same order, $\gamma$ is
usually a little larger), but also the most tedious to calculate. To
do this we begin by considering the 2 interaction terms
\begin{equation}
V_{12}^{S \tau} = -\sum_{\alpha,\beta} \gamma_1^{\alpha \beta}
\frac{\partial X_{1 \alpha}}{\partial x_{1 \beta}} S_1^z -
\sum_{\gamma,\delta,\eta} \zeta_2^{\gamma \delta \eta}
\frac{\partial^2 X_{2 \gamma}}{\partial x_{2 \delta} \partial x_{2
\eta}} \tau_2^z .
\end{equation}
which come into the calculation of this interaction when we have
only 2 impurities. Using Eq.(\ref{Fourier}) we obtain
\begin{eqnarray}
V_{12}^{S \tau}=-\frac{1}{\sqrt{N}} \sum_{q \mu} \sum_{\alpha \beta}
\gamma_1^{\alpha \beta} X_{q \mu} \bfe_{q \mu \alpha} i q_{\beta}
e^{i q x_1} S_1^z \nonumber \\ +\frac{1}{\sqrt{N}} \sum_{q \mu}
\sum_{\gamma \delta \eta} \zeta_2^{\gamma \delta \eta} X_{q \mu}
\bfe_{q \mu \gamma} q_{\delta} q_{\eta} e^{i q x_2} \tau_2^z .
\end{eqnarray}
Again, minimizing the sum of $V_{S \tau}$ and the potential term in
Eq.(\ref{phonon}) one obtains the interaction energy between the two
impurities, given by:
\begin{equation}
U_{12}^{S \tau} = - \sum_{q \mu} \frac{ \sum_{\alpha \beta \gamma
\delta \eta} \gamma_1^{\alpha \beta} \zeta_2^{\gamma \delta \eta}
\bfe_{q \mu \alpha} \bfe_{q \mu \gamma} q_{\beta} q_{\delta}
q_{\eta} \sin{(q \cdot R)} S_1^z \tau_2^z}{NM \omega_{q \mu}^2} .
\end{equation}

Notice that this interaction is odd in ${\bf R}$! This is because
the first derivative term is imaginary, and the second derivative
term is real, leading to a $\sin{({\bf q \cdot R})}$ interaction.
Physically, this is clear since an impurity variable which is odd
under inversion symmetry, like $\tau_2$, must have an interaction
which is odd in ${\bf R}$ with other impurities, if these are either
substitutional (ie., causing scalar perturbations), or are even
under inversion.

Using again the acoustic approximation, i.e. $\omega_{q l} = c_l q,
\omega_{q \perp} = c_{\perp} q$ and the identities in (\ref{polI}),
we have an interaction
\begin{eqnarray}
U_{12}^{S \tau} &=& -\frac{1}{NM} \sum_q \frac{1}{q^2} \sum_{\alpha
\beta \gamma \delta \eta} \gamma_{1}^{\alpha \beta} \zeta_2^{\gamma
\delta \eta} q_{\beta} q_{\delta} q_{\eta} \sin{(q \cdot R)} S_1^z
\tau_2^z \nonumber \\ & \times & \left[ \left( \frac{1}{c_l^2} -
\frac{1}{c_{\perp}^2} \right) \frac{q_{\alpha} q_{\gamma}}{q^2} +
\frac{1}{c_{\perp}^2} \delta_{\alpha \gamma} \right]
 \label{U-St}
\end{eqnarray}
This is a complicated integral, because of the large number of
different components of momentum involved. To evaluate it we begin
by writing it in the form:
\begin{eqnarray}
U_{12}^{S \tau} &=&  -\frac{1}{NM} \sum_{\alpha \beta \gamma \delta
\eta} \gamma_1^{\alpha \beta} \zeta_2^{\gamma \delta \eta} S_1^z
\tau_2^z \nonumber \\ & \times & \left[ \left( \frac{1}{c_l^2} -
\frac{1}{c_{\perp}^2} \right) {\cal F}_{\alpha \beta \gamma \delta
\eta} \;+\; \frac{1}{c_{\perp}^2} \delta_{\alpha \gamma} \; {\cal
G}_{\beta \delta \eta}  \right],
 \label{U-St2}
\end{eqnarray}
where we have define the 5th and 3rd rank tensors
\begin{equation}
{\cal F}_{\alpha \beta \gamma \delta \eta} = \sum_q \frac{q_{\alpha}
q_{\beta} q_{\gamma} q_{\delta} q_{\eta} \sin{({\bf q} \cdot {\bf
R})}}{q^4}
\end{equation}

\begin{equation}
{\cal G}_{\beta \delta \eta} = \sum_q \frac{q_{\beta} q_{\delta}
q_{\eta} \sin{({\bf q} \cdot {\bf R})}}{q^2} .
\end{equation}

Let us start by calculating ${\cal G}_{\beta \delta \eta}$. As a
first step we write it in the form
\begin{equation}
{\cal G}_{\beta \delta \eta} = \frac{V}{(2 \pi)^3} \frac{1}{R^4}
G_{\beta \delta \eta} .
\end{equation}
where we have defined the integral
\begin{equation}
G_{\beta \delta \eta} \equiv \int d^3q \frac{q_{\beta} q_{\delta}
q_{\eta} \sin{(q \cdot \bfn)}}{q^2} .
\end{equation}
Already here we see that the interaction has a spatial dependence
$\sim 1/R^4$. In fact, any interaction involving the $\tau_z$
impurity will have a $1/R^4$ or larger power spatial dependence (we
find that the $\tau^z \tau^z$ impurity-impurity interaction has a
$1/R^5$ dependence), and must therefore be treated as a short range
interaction in $3$D.

By symmetry the function $G_{\beta \delta \eta}$ can be written in
the form
\begin{equation}
G_{\beta \delta \eta} = A ( \delta_{\beta \delta} \bfn_{\eta} +
\delta_{\beta \eta} \bfn_{\delta} + \delta_{\delta \eta}
\bfn_{\beta} ) + B \bfn_{\beta} \bfn_{\delta} \bfn_{\eta} .
\end{equation}

We wish to find the coefficients $A$ and $B$. We do this along the
same lines as in our calculation for the random field term of the
$S^z$ impurities. First we note that
\begin{equation}
\sum_{\alpha \eta} G_{\alpha \alpha \eta} \bfn_{\eta} = \sum_{\eta}
\int d^3q q_{\eta} \bfn_{\eta} \sin{(q \cdot \bfn)}
\end{equation}
is a scalar. Thus the integral can be taken for ${\bf }\eta =
\hat{z}$, and is zero. On the other hand the sum on the left side is
$5A+B$, and therefore we get
\begin{equation}
5A + B = 0 .
\end{equation}

We now consider
\begin{equation}
\sum_{\beta \delta \eta} G_{\beta \delta \eta} \bfn_{\beta}
\bfn_{\delta} \bfn_{\eta} = \sum_{\beta \delta \eta} \int d^3q
\frac{q_{\beta} q_{\delta} q_{\eta} \bfn_{\beta} \bfn_{\delta}
\bfn_{\eta}}{q^2} \sin{(q \cdot \bfn)} .
\end{equation}
The left side equals $3A+B$. On the other hand, since the expression
is a scalar the integral can be calculated for $\bfn = n_z$, i.e.
\begin{eqnarray}
& & \int d^3q \frac{q_z^3 \sin{(q_z)}}{q^2} \nonumber \\ &=& 2 \pi
\int_0^{\infty} dq_r q_r \int_{- \infty}^{\infty} dq_z q_z
\sin{(q_z)} \left(1 - \frac{q_r^2}{q_z^2 + q_r^2} \right) \nonumber
\\ &=& -12 \pi^2 .
\end{eqnarray}
Thus we find that $3A+B = -12 \pi^2$, and since $5A+B=0$ we find
that $A=6 \pi^2$, $B=-30 \pi^2$, and therefore finally we have
\begin{equation}
G_{\beta \delta \eta} = \pi^2 [6 ( \delta_{\beta \delta} \bfn_{\eta}
+ \delta_{\beta \eta} \bfn_{\delta} + \delta_{\delta \eta}
\bfn_{\beta} ) - 30 \bfn_{\beta} \bfn_{\delta} \bfn_{\eta}]
 \label{Gbdn}
\end{equation}
and hence
\begin{equation}
{\cal G}_{\beta \delta \eta} = \frac{V}{4 \pi R^4} [3 (
\delta_{\beta \delta} \bfn_{\eta} + \delta_{\beta \eta}
\bfn_{\delta} + \delta_{\delta \eta} \bfn_{\beta} ) - 15
\bfn_{\beta} \bfn_{\delta} \bfn_{\eta}] .
 \label{Gbdn2}
\end{equation}

Let us now calculate ${\cal F}_{\alpha \beta \gamma \delta \eta}$.
We first define
\begin{equation}
{\cal F}_{\alpha \beta \gamma \delta \eta} = \frac{V}{(2 \pi)^3}
\frac{1}{R^4} F_{\alpha \beta \gamma \delta \eta}
\end{equation}
where
\begin{equation}
F_{\alpha \beta \gamma \delta \eta} = \int d^3q \frac{q_{\alpha}
q_{\beta} q_{\gamma} q_{\delta} q_{\eta} \sin{(q \cdot \bfn)}}{q^4}
.
\end{equation}

From symmetry, one can write:
\begin{equation}
F_{\alpha \beta \gamma \delta \eta} = a F^a_{\alpha \beta \gamma
\delta \eta} + b F^b_{\alpha \beta \gamma \delta \eta} + c
F^c_{\alpha \beta \gamma \delta \eta}
 \label{Fabc}
\end{equation}
where
\begin{eqnarray}
F^a_{\alpha \beta \gamma \delta \eta} &=& \delta_{\alpha \beta}
\delta_{\gamma \delta} \bfn_{\eta} + \delta_{\alpha \gamma}
\delta_{\beta \delta} \bfn_{\eta} + \delta_{\alpha \delta}
\delta_{\beta \gamma} \bfn_{\eta} \nonumber \\ &+& (\eta
\leftrightarrow \alpha + \eta \leftrightarrow \beta + \eta
\leftrightarrow \gamma + \eta \leftrightarrow \delta)    \nonumber
\\ F^b_{\alpha \beta \gamma \delta \eta} &=& \delta_{\alpha \beta}
\bfn_{\gamma} \bfn_{\delta} \bfn_{\eta} \;+\; {\rm all} \; {\rm
distinguishable} \; {\rm permutations}    \nonumber \\
F^c_{\alpha \beta \gamma \delta \eta} &=& \bfn_{\alpha} \bfn_{\beta}
\bfn_{\gamma} \bfn_{\delta} \bfn_{\eta}.
 \label{abc}
\end{eqnarray}

We now need to evaluate the three coefficients $a,b$, and $c$. To do
so let us first consider the term
\begin{equation}
\sum_{\alpha \beta \eta} F_{\alpha \alpha \beta \beta \eta}
\bfn_{\eta} = \int d^3q (q \cdot \bfn) \sin{(q \cdot \bfn)} .
\end{equation}
The equality results because the sum gives $q^4 q_{\eta}$ in the
numerator. This expression is a scalar, which therefore equals $\int
d^3q q_z \sin{q_z} = 0$. A careful evaluation of the sum on the left
side  gives $35a + 14b + c$ and therefore we obtain as a first
relation for $a,b,c$ that
\begin{equation}
35a + 14b + c =0 .
\end{equation}

To obtain a second relation we look at the scalar:
\begin{eqnarray}
& & \sum_{\alpha \beta \gamma \delta \eta} F_{\alpha \beta \gamma
\delta \eta} \delta_{\alpha \beta} \bfn_{\gamma} \bfn_{\delta}
\bfn_{\eta} \nonumber \\ &=& \sum_{\gamma \delta \eta} \int d^3q
\frac{q_{\gamma} q_{\delta} q_{\eta} \bfn_{\gamma} \bfn_{\delta}
\bfn_{\eta}}{q^2} \sin{(q \cdot \bfn)}
\end{eqnarray}
Since the expression is a scalar, we can choose $\bfn = n_z$, and
the integral becomes $\int d^3q q_z^3 \sin{q_z}/q^2 = -12 \pi^2$, as
was calculated above. Summation of the left side gives $21a + 12b +
c$, and we thus have a second relation between $a,b$, and $c$, viz.:
\begin{equation}
21a + 12b +c = -12 \pi^2 .
\end{equation}

To get a third relation we consider
\begin{equation}
\sum_{\alpha \beta \gamma \delta \eta} F_{\alpha \beta \gamma \delta
\eta} \bfn_{\alpha} \bfn_{\beta} \bfn_{\gamma} \bfn_{\delta}
\bfn_{\eta} = \int d^3q q_z^5 \sin{q_z}/q^4 ,
\end{equation}
where we have used, as above, the fact that the expression is a
scalar and took $\bfn = n_z$. The integral can be calculated using
again the identity $q_z^2/q^2 = 1 - q_r^2/q^2$, and is found to
equal zero. The sum on the left side then gives the relation
\begin{equation}
12a + 10b + c = 0
\end{equation}
If we now take these three relations together, we find the desired
results for $a,b$ and $c$:
\begin{eqnarray}
a = 24 \pi^2/5  \nonumber \\
b = -138 \pi^2/5  \nonumber \\
c = 1092 \pi^2/5 .
   \label{abcR}
\end{eqnarray}
which can then be inserted into Eq.(\ref{Fabc}) to get a final
result for the interaction $U_{12}^{S \tau}$ in the form given in
(\ref{U-St2}).

Now we can write the form of the effective random field term that
this leads to, after noting again that almost all the $\{ {\bf S}_j
\}$ are frozen, and sum over all sites apart from a given site $i$.
Then we must get a term in the low-$T$ effective Hamiltonian of form
\begin{equation}
H_{(\gamma)}^{\tau} \;=\; \sum_i b_i^{(\gamma)} \hat{\tau}_i^z
 \label{Htau1}
\end{equation}
where the random field $b_i^{(\gamma)}$ is given by summing over
sites in the interaction we have just derived. One gets
\begin{eqnarray}
b_i^{(\gamma)}  &=&  -\frac{V}{2 \pi^3 NM} \sum_j {1 \over R_{ij}^4}
\sum_{\alpha \beta \gamma \delta \eta} \gamma_j^{\alpha \beta}
\zeta_i^{\gamma \delta \eta} \langle S_j^z \rangle \nonumber
\\ & \times & \left[ \left( \frac{1}{c_l^2} - \frac{1}{c_{\perp}^2} \right)
F^{ij}_{\alpha \beta \gamma \delta \eta} \;+\; \frac{2
\pi^2}{c_{\perp}^2} \delta_{\alpha \gamma} \; G^{ij}_{\beta \delta
\eta} \right],
 \label{Ub-St2}
\end{eqnarray}
where the tensor function $F^{ij}_{\alpha \beta \gamma \delta \eta}$
is just
\begin{eqnarray}
\hspace{-3mm} & & F^{ij}_{\alpha \beta \gamma \delta \eta} = \int
d^3q \frac{q_{\alpha} q_{\beta} q_{\gamma} q_{\delta} q_{\eta}
\sin{({\bf q} \cdot {\bf n}_{ij})}}{q^4} \nonumber \\ \hspace{-3mm}
&=& [a F^a_{\alpha \beta \gamma \delta \eta}({\bf n}_{ij}) + b
F^b_{\alpha \beta \gamma \delta \eta}({\bf n}_{ij}) + c F^c_{\alpha
\beta \gamma \delta \eta}({\bf n}_{ij})]
 \label{Fijabc}
\end{eqnarray}
with the 3 tensors in the 2nd form given by substituting the unit
${\bf n}_{ij} = R_{ij}/\vert {\bf R}_{ij} \vert$ for ${\bf n}$ in
(\ref{abc}); and the tensor $G^{ij}_{\beta \delta \eta}$ is given by
the form in (\ref{Gbdn}) after the same substitution has been made,
ie..
\begin{equation}
G_{ij}^{\beta \delta \eta} = \pi^2 [6 ( \delta^{\beta \delta}
\bfn_{ij}^{\eta} + \delta^{\beta \eta} \bfn_{ij}^{\delta} +
\delta^{\delta \eta} \bfn_{ij}^{\beta} ) - 30 \bfn_{ij}^{\beta}
\bfn_{ij}^{\delta} \bfn_{ij}^{\eta}] .
 \label{Gbdnij}
\end{equation}

This is the first random field term acting on the $\{ \hat{\tau}_j
\}$. We see that the order of magnitude of the interaction is given
by
\begin{equation}
\vert b_i^{(\gamma)} \vert \sim \frac{\gamma \zeta}{\rho c^2} \sum_j
{1 \over R_{ij}^4} .
 \label{gammazeta}
\end{equation}
where $\gamma$ and $\zeta$ are typical values of the corresponding
tensors.

%%%%%%%%%%%%%%%%%%%%%%%%%%%%%%%%%%%%%%%%%%%%%%%%%%%%%%%
\subsubsection{The volume interaction term}
 \label{sec:etatau}
%%%%%%%%%%%%%%%%%%%%%%%%%%%%%%%%%%%%%%%%%%%%%%%%%%%%%%%

Now let us consider the other source of random fields acting on the
$\{ \hat{\tau}_j \}$, coming from their interaction with the volume
term of other impurities. This is calculated by considering the
cross-term arising from the interaction
\begin{equation}
V_{\eta \tau} = -\sum_{\alpha} \eta_1 \frac{\partial X_{1
\alpha}}{\partial x_{1 \alpha}} - \sum_{\gamma,\delta,\eta}
\zeta_2^{\gamma \delta \eta} \frac{\partial^2 X_{2 \gamma}}{\partial
x_{2 \delta} \partial x_{2 \eta}} \tau_2^z .
\end{equation}
between the volume interaction at site ${\bf r}_1$ and
$\hat{\tau}_2^z$. Similar manouevres to the ones used above then
lead, in the acoustic approximation, to an interaction term
\begin{eqnarray}
U_{\eta \tau} &=& \frac{-1}{NM} \sum_q \frac{1}{q^2} \sum_{\alpha
\gamma \delta \eta} \eta_1 \zeta_2^{\gamma \delta \eta} q_{\alpha}
q_{\delta} q_{\eta} \sin{({\bf q} \cdot {\bf R})} \tau_2^z \nonumber
\\ & \times & \left[ \left( \frac{1}{c_l^2} - \frac{1}{c_{\perp}^2} \right)
\frac{q_{\alpha} q_{\gamma}}{q^2} + \frac{1}{c_{\perp}^2}
\delta_{\alpha \gamma} \right] ,
 \label{U-etaT}
\end{eqnarray}
which reduces to
\begin{equation}
U_{\eta \tau} \;=\; \frac{-1}{NM c_l^2} {\cal G}_{\gamma \delta
\eta} \eta_1 \zeta_2^{\gamma \delta \eta} \tau_2^z  \;\;\equiv\;\;
\frac{-1}{8 \pi^3 \rho R^4 c_l^2} G_{\gamma \delta \eta} \eta_1
\zeta_2^{\gamma \delta \eta} \tau_2^z ,
\end{equation}
where the definitions of ${\cal G}$ and $G$ are the same as above
[{\it cf.} Eqs. (\ref{Gbdn}) and (\ref{Gbdn2})]. Note that since the
volume term couples only to the longitudinal phonons, $U_{\eta
\tau}$ depends only on $c_l$.

The random field resulting from this term is then given in the form
of an interaction
\begin{equation}
H_{(\eta)}^{\tau} \;=\; \sum_i b_i^{(\eta)} \hat{\tau}_i^z
 \label{Htau2}
\end{equation}
where the random field $b_i^{(\eta)}$ is given by
\begin{equation}
b_i^{(\eta)} \;=\; -\frac{1}{8 \pi^3 \rho c_l^2}\sum_j {1 \over
R_{ij}^4} G^{ij}_{\gamma \delta \eta} \eta \zeta_i^{\gamma \delta
\eta}
 \label{btau2}
\end{equation}
with $G^{ij}_{\gamma \delta \eta}$ given by (\ref{Gbdnij}) above,
and where
\begin{equation}
\vert b_i^{(\eta)} \vert \sim \frac{\eta \zeta}{\rho c^2} \sum_j {1
\over R_{ij}^4} .
 \label{gammaeta}
\end{equation}

Now we can finally write the random fields acting on the $\{
\hat{\tau}_j \}$, coming from the gradient phonon term in the
defect-phonon interaction, in the form of an interaction term
\begin{equation}
H_{RF}^{\tau} \;=\; \sum_i b_i \hat{\tau}_i^z
 \label{HtauR}
\end{equation}
where the random field $b_i$ is just given by the sum of the two
contribution we have found, ie.,
\begin{equation}
b_i \;=\; b_i^{(\eta)} + b_i^{(\gamma)}
 \label{biF}
\end{equation}
where the two contributions are given by eqtns. (\ref{U-St2}) and
(\ref{btau2}) respectively, and where $\vert
b_i^{(\gamma)}/b_i^{(\eta)} \vert \sim \gamma/\eta$.

%%%%%%%%%%%%%%%%%%%%%%%%%%%%%%%%%%%%%%%%%%%%%%%%%%%%%%%%%%%%%%%%%%%%%%%%%%%%%%%%%%%%%%%%%%
\subsubsection{Ising interaction between the $\{ \hat{\tau}_j
\}$}
 \label{sec:otherI}
%%%%%%%%%%%%%%%%%%%%%%%%%%%%%%%%%%%%%%%%%%%%%%%%%%%%%%%%%%%%%%%%%%%%%%%%%%%%%%%%%%%%%%%%%%

In the same way we may evaluate the Ising interaction between the
$\{ \hat{\tau} \}$.  The result is an interaction $\sum_{ij}J_{ij}
\hat{\tau}_i^z \hat{\tau}_j^z$, where the interaction coefficient
\begin{equation}
J_{ij}  \;\sim\; \frac{\zeta^2}{\rho c_o^2}  {1 \over R_{ij}^5} .
 \label{zeta2}
\end{equation}
Now we see that not only is the interaction coefficient here much
smaller than that for the random fields (since $\zeta \ll \eta,
\gamma$), but this interaction also falls off much faster with
$R_{ij}$, going like $1/R_{ij}^5$.  For this reason we do not
calculate the exact coefficient $J_{ij}$ here: this calculation is
rather lengthy, and it simply multiplies the right hand of
(\ref{zeta2}) by a complicated angular factor $\sim O(1)$. Note that
even though $J_{ij}$ is so small, it will still affect the dynamics
of the system at low $T$, since there will be still resonant $\{
\hat{\tau}_j \}$ that can tunnel.

%%%%%%%%%%%%%%%%%%%%%%%%%%%%%%%%%%%%%%%%%%%%%%%%%%%%%%%%%%%%%%%%%%%%%%%%%%%%%%%%%%%%%%%%%%
\subsection{Effective Hamiltonian: low-Energy Form}
 \label{sec:Heff-F}
%%%%%%%%%%%%%%%%%%%%%%%%%%%%%%%%%%%%%%%%%%%%%%%%%%%%%%%%%%%%%%%%%%%%%%%%%%%%%%%%%%%%%%%%%%

Let us now summarize what we have. After integrating out the
phonons, we can now say that we have ended up with an effective
Hamiltonian which, if we still treat all the variables as operators,
takes the form
\begin{equation}
H_{eff} \;=\; - \sum_j [D_j \hat{S}_j^x \;+\; \Delta_j
\hat{\tau}_j^x]  \;\;+\;\; V_{eff}
  \label{Htotal}
\end{equation}
where the interaction now contains the following terms:
\begin{eqnarray}
V_{eff} &=& \sum_{ij} U_{ij}^{zz} S_i^z S_j^z + \sum_i B_i S_i^z +
\sum_{ij} U^{S\tau}_{ij} S_i^z \tau_j^z \nonumber \\ &+& \sum_i
b_i^{(\eta)} \tau_i^z + \sum_{ij} J_{ij} \tau_i^z \tau_j^z ,
\end{eqnarray}
where we have written the terms in decreasing order of their
strength. The 2 random fields in this effective Hamiltonian arise
from the phonon-mediated coupling of the $\{ \hat{S}_i^z \}$ and the
$\{ \hat{\tau}_i^z \}$ variables to the volume distortion caused by
the defects.

The above effective Hamiltonian must be used if we want to analyse
the dynamics of these variables. At high energies $\sim T_G$, we can
entirely drop all the terms involving the $\{ \hat{\tau}_i^z \}$ in
the interaction $V_{eff}$, since these interactions are all $\sim
O(\zeta)$ and too weak to play a role. If we ignore the small number
of spins $S_j$ that are in resonance, then we can also assume that
the $\{ S_j^z \}$ variables are frozen by the strong Ising
interaction $U_{ij}^{zz}$. As discussed elsewhere, the effect of the
random field $B_i$ is then to destroy long-range glassy
order\cite{SS06}.

Now suppose we go to low energy scales. If we continue to ignore the
small quantum fluctuations in the expectation values $\{ \langle
\hat{S}_j \rangle \}$ brought about by tunneling of the small
concentration of resonant $S_j$, then we can treat this distribution
as frozen. Then the low-energy Hamiltonian simplifies very
considerably. We get
\begin{equation}
H_{eff} \;\rightarrow \;\; - \sum_j \Delta_j \hat{\tau}_j^x
\;\;+\;\; V_{eff}
 \label{HeffL}
\end{equation}
where now the interaction term has the much simpler form
\begin{equation}
V_{eff} \;\rightarrow \; \sum_i b_i \hat{\tau}_i^z \;+\; \sum_{ij}
J_{ij} \hat{\tau}_i^z \hat{\tau}_j^z
 \label{V-lowT}
\end{equation}
in which the random field $b_i$ was calculated in the last section,
and, as noted before, it is much larger than the Ising interaction
$J_{ij}$.

%%%%%%%%%%%%%%%%%%%%%%%%%%%%%%%%%%%%%%%%%%%%%%%%%%%%%%%%%%%%%%%%%%%%%%%%%%%%%%%%%%%%%%%%%%%%%%%%%%%
%%%%%%%%%%%%%%%%%%%%%%%%%%%%%%%%%%%%%%%%%%%%%%%%%%%%%%%%%%%%%%%%%%%%%%%%%%%%%%%%%%%%%%%%%%%%%%%%%%%
\section{Summary and Remarks}
 \label{sec:sum}
%%%%%%%%%%%%%%%%%%%%%%%%%%%%%%%%%%%%%%%%%%%%%%%%%%%%%%%%%%%%%%%%%%%%%%%%%%%%%%%%%%%%%%%%%%%%%%%%%%%
%%%%%%%%%%%%%%%%%%%%%%%%%%%%%%%%%%%%%%%%%%%%%%%%%%%%%%%%%%%%%%%%%%%%%%%%%%%%%%%%%%%%%%%%%%%%%%%%%%%

The purpose of the present paper was to give a detailed treatment of
the phonon-mediated interactions which exist in a neutral glass,
taking into account not only the usual linear coupling between
defects and the phonon displacement field, but also the coupling to
the gradient of the phonon field. The net result of this was that we
found an effective Hamiltonian for the two sets of variables (the
rotational tunneling variables $\{ \hat{S}_j \}$ and the inversion
tunneling variables $\{ \hat{\tau}_j \}$), which contained tunneling
terms for each, Ising interaction terms for each, and various
effective random fields which act on both variables. We note that
similar issues arise in quantum spin glasses\cite{TH95,SS05,SL06},
where random field terms also have a profound effect (just as they
do in classical spin glasses\cite{FH86}); however real spins are in
many ways quite different from defects, and this means that there is
no simple relation between the two systems\cite{SS06}.

These calculations were all done in the framework of 2nd-order
perturbation theory in the interactions, and they were all done
assuming that the background lattice could still be meaningfully
defined, at least in local "patches" around each defect. Thus at
first glance the calculations here are only rigourously valid for
low defect concentration $x$. Two possible problems then arise at
higher $x$. First, one might object that the 'patch' picture must
eventually break down - we have argued in section \ref{sec:bareH}
that this is not the case, because even rather strongly disordered
glasses still do have local crystalline order.

The second more serious problem is that one expects higher-order
interactions to come in at higher $x$ and these will mix the various
interactions we have derived here. This problem of higher-order
corrections is notoriously difficult, since a hierarchy of logs is
generated once one integrates over multiple
sites\cite{BNOK98,Lev90}. We do not attempt to discuss it here, but
simply note that our results inevitably changes the results of these
higher-order calculations, because of the new terms we have found in
this paper (in fact almost all calculations of these higher order
terms include only the tunneling terms and the Ising interactions,
without any random fields). Thus we expect the results here to have
important consequences for the discussion of the nature of glasses,
and we have developed some of these elsewhere\cite{SS06}.

%%%%%%%%%%%%%%%%%%%%%%%%%%%%%%%%%%%%%%%%%%%%%%%%%%%%%%%%%%%%%%%%%%%%%%%%%%%%%%%%%%%%%%%%
%%%%%%%%%%%%%%%%%%%%%%%%%%%%%%%%%%%%%%%%%%%%%%%%%%%%%%%%%%%%%%%%%%%%%%%%%%%%%%%%%%%%%%%%
\section{Acknowledgements}
 \label{sec:Thanks}
%%%%%%%%%%%%%%%%%%%%%%%%%%%%%%%%%%%%%%%%%%%%%%%%%%%%%%%%%%%%%%%%%%%%%%%%%%%%%%%%%%%%%%%%
%%%%%%%%%%%%%%%%%%%%%%%%%%%%%%%%%%%%%%%%%%%%%%%%%%%%%%%%%%%%%%%%%%%%%%%%%%%%%%%%%%%%%%%%

We would like to thank A. Burin, J. Rottler, G.A. Sawatzky, B.
Seradjeh, and A.P. Young for very useful discussions. This work was
supported by NSERC of Canada, by the Canadian Institute for Advanced
Research, and by the Pacific Institute of Theoretical Physics.

\end{document}